\documentclass[aps,amsmath,twocolumn,floatfix,superscriptaddress,footinbib]{revtex4-1}
\pdfoutput=1
\usepackage[dvips]{graphics}
\usepackage{bm}
\usepackage{import}
\usepackage{epsfig}
\usepackage{enumerate}
\usepackage{subfigure}
\usepackage{color}
\usepackage{xcolor}
\usepackage{epsfig,amssymb,amsmath,latexsym}
\usepackage{sidecap,tikz}
\usepackage{graphicx}
\usepackage{makecell}
\usepackage{hyperref}
\usepackage{multirow}
\usepackage[normalem]{ulem}

\hypersetup{
    pdfnewwindow=true,      % links in new window
    colorlinks=true,       % false: boxed links; true: colored links
    linkcolor=blue,          % color of internal links
    citecolor=magenta,        % color of links to bibliography
    filecolor=blue,      % color of file links
    urlcolor=blue        % color of external links
}

\newcommand{\ie}{{i.e., }}

\newcommand{\bea}{\begin{eqnarray}}
\newcommand{\eea}{\end{eqnarray}}
\newcommand{\beq}{\begin{equation}}  
\newcommand{\eeq}{\end{equation}}
\newcommand{\non}{\nonumber}

\definecolor{lime}{HTML}{A6CE39}
\DeclareRobustCommand{\orcidicon}{\hspace{-1mm}
	\begin{tikzpicture}
	\draw[lime, fill=lime] (0,0) 
	circle [radius=0.16] 
	node[white] {{\fontfamily{qag}\selectfont \tiny \,ID}};
	\draw[white, fill=white] (-0.0525,0.095) 
	circle [radius=0.007];
	\end{tikzpicture}
	\hspace{-3mm}
}
\foreach \x in {A, ..., Z}{\expandafter\xdef\csname orcid\x\endcsname{\noexpand\href{https://orcid.org/\csname orcidauthor\x\endcsname}
			{\noexpand\orcidicon}}
}

\begin{document} 
\title{Phase-dependent charge and heat current in thermally biased short Josephson junctions formed at helical edge states}
\author{Paramita Dutta\orcidA{}}
\affiliation{Theoretical Physics Division, Physical Research Laboratory, Navrangpura, Ahmedabad-380009, India}
\email{paramita@prl.res.in}
\date \today

\begin{abstract}

We explore the phase-dependent charge and heat current in the short Josephson junctions with two normal metal regions attached at opposite ends, formed at helical edge states of two-dimensional topological insulators (TIs). For all finite phases, an asymmetry appears around the zero energy in the transmission spectra except for $\phi\!=\!n\phi_0$, where $n$ is a half-integer and $\phi_0$  ($=2\pi$) is the flux quantum. The phase-induced asymmetry plays a key role in inducing charge and heat current through the thermally biased junction. However, the current amplitudes are sensitive to the size of the junction. We show that in the short Josephson junction when subject to a temperature gradient, the charge current shows an odd-symmetry in phase. It indicates that the phase-tunable asymmetry around the zero-energy is not sufficient to induce a dissipative thermoelectric current in the junction. This is in contrast to the behavior of long Josephson junction as shown in the literature. The phase-tunable heat currents are obtained with amplitudes set by the phase difference, base temperature, and system size.

\end{abstract}

\maketitle

%------------------------------------------------- %
\section{Introduction}
 % -------------------------------------------------- %
 
Study of thermal gradient-induced current in superconductors and superconducting junctions has been rejuvenated in recent years, breaking the concepts of poor thermal current in superconductors\,\cite{Ozaeta2014,Kolenda2016,Kolenda2016b,Linder2016,Dutta2017,Beckmannprb2019,Dutta2020b,Ouassou2022}. The thermal current in ordinary superconductors were expected to be low or even vanishing, primarily because of the superconducting gap in the density of states. The symmetry in the energy spectrum is responsible for low charge current in the linear regime\,\cite{Giazotto2006}. On top of that, thermal bias-induced charge current interferes with the superflow, and this causes the separation of the two currents tricky. For these reasons, conventional Bardeen-Cooper-Schrieffer (BCS) superconductors were not considered as active thermoelectric materials for several years\,\cite{Galperin2002}. Unconventional superconductors were also studied in few works to enhance the thermal current \ie the non-dissiptaive charge current\,\cite{Tomas2004,Seja2022}.
 
Recently, some efforts have been put to enhance thermal charge current, within the linear regime, in superconducting junctions instead of bare superconductors by breaking the spin-symmetry using ferromagnetic elements\,\cite{Ozaeta2014}, forming ferromagnet/superconductor\,\cite{Machon2014,Kolenda2016,Kolenda2016b,Linder2016,Dutta2017,Beckmannprb2019,Dutta2020b,Ouassou2022} or anti-ferromagnet/superconductor hybrid structures\,\cite{Jakobsen2020}. Research in this direction has been boosted after the experimental verification in 2016\,\cite{Kolenda2016}, where an excellent agreement with the theoretical prediction\,\cite{Ozaeta2014} was confirmed. Very recently. it has been predicted that a nonlinear thermal current can flow in the presence of spontaneously broken particle-hole symmetry\,\cite{Marchegiani2020}.

Search for ways to control the thermal currents in superconductor junctions is continued. In recent work, Kalenkov et al. have shown that depending on the topology and the temperature gradient it is possible to generate a large phase-coherent charge current in a Josephson junction (JJ)\,\cite{Kalenkov2020}. In JJ, one can avoid using external elements like non-magnetic or magnetic impurity\,\cite{Kalenkov2012}, or any engineering like creating vacancy\,\cite{Aydin2022}, which has been utilized to enhance thermal current in other junctions. In fact, non-trivial thermal bias-induced voltage can be achieved just by tuning the superconducting phase of JJ\,\cite{Guttman1997,Kalenkov2020,Marchegiani2020,Blasi2020a,Blasi2020b}.
In another work, phase-tunable thermal-bias induced charge current is shown in a ballistic junction\,\cite{Mukhopadhyay2022}. Hence, the phase-tunability makes JJ more powerful compared to other superconducting junctions.

To generate the phase-tunable thermal current in JJ, topological materials have also been considered in very few works\,\cite{Blasi2020a,Blasi2020b,Blasi2021,Gresta2021,Saxena2022,Mukhopadhyay2022} as the combination of global topology and local superconducting order has been established to host exotic transport properties in the literature\,\cite{Black-Schaffer2012,Black-Schaffer2013,Cayao2017,Dutta2019,Dutta2020a}. Particularly, junctions involving two-dimensional ($2$D) topological insulators (TI)\,\cite{Kane2005} have drawn great attention because of its potential to influence scattering processes\,\cite{Tanaka2004,Rolf2013,Islam2017} and most importantly to host Majorana fermions\,\cite{Kane2005,Fu2008,Fu2009,Qi2011,Tanaka2004,Tkachov2013}. The one-dimensional ($1$D) helical edge states make $2$D\,TIs\,\cite{Hasan2010} more effective by preventing all the backscatterings but admitting only two processes:\,(i) Andreev reflections and (ii) electron transmissions through the junction\,\cite{Cayao2017,Cayao2022}. Also, there are recent predictions for the  detection of topological bound states via thermal current in some junctions including JJ\,\cite{Savander2020,Mukhopadhyay2022}. 

Notably, the charge current consists of dissipative and non-dissipative parts. The dissipative charge current describes the conventional thermoelectricity while, the traditional non-dissipative Josephson current is also unavoidable in the same junction. The previous studies involve JJ where the superconductors act as leads with various widths of the middle normal regions. The effect of finite sized superconductors is yet to explore. Motivated by this, we study the charge and heat current in thermally biased short JJ (sJJ) when it is formed at the $1$D helical edges of $2$D\,TI in proximity to ordinary superconductor. Within the short JJ, an extremely short normal region is sandwiched between two finite size superconductors. The phase-tunable topological Andreev bound states (ABS) formed in such normal metal/short Josephson junction/normal metal (N-sJJ-N) junction at the edge of $2$D\,TI help generate charge current under voltage bias condition as seen in Ref.\,[\onlinecite{Cayao2022}] which further motivates to search for the phase-tunable thermal current in the same junction. We show that the appearance of the asymmetry around the zero energy in the transmission spectra plays a key role here. The charge and heat currents generated by the thermal gradient are tunable by the phase of the junction with their amplitudes being sensitive to the lengths of the finite-sized superconductors. Remarkably, the charge current generated in the junction is entirely non-dissipative. The phase-tunability is not sufficient to produce a dissipative charge current when the junction is short, which is in contrast to the previous results in long JJ. We demonstrate that the charge and heat current can be optimized when the superconductors' lengths are of the order of coherence length.  Our work thus predicts topological Josephson junction, where the heat and dissipative charge current is smoothly controllable by the phase of the junction, can behave in a way completely different from the long JJ and thus help in choosing proper system size for thermal current.

%----------------------------------------------
\section{Model and Hamiltonian}\label{model}
%----------------------------------------------
 %%%%%%%%%%%%%%%%%%%% FIGURE OF BAND STRUCTURE %%%%%%%%%%%
\begin{figure}[!thpb]
\centering
\includegraphics[scale=.19]{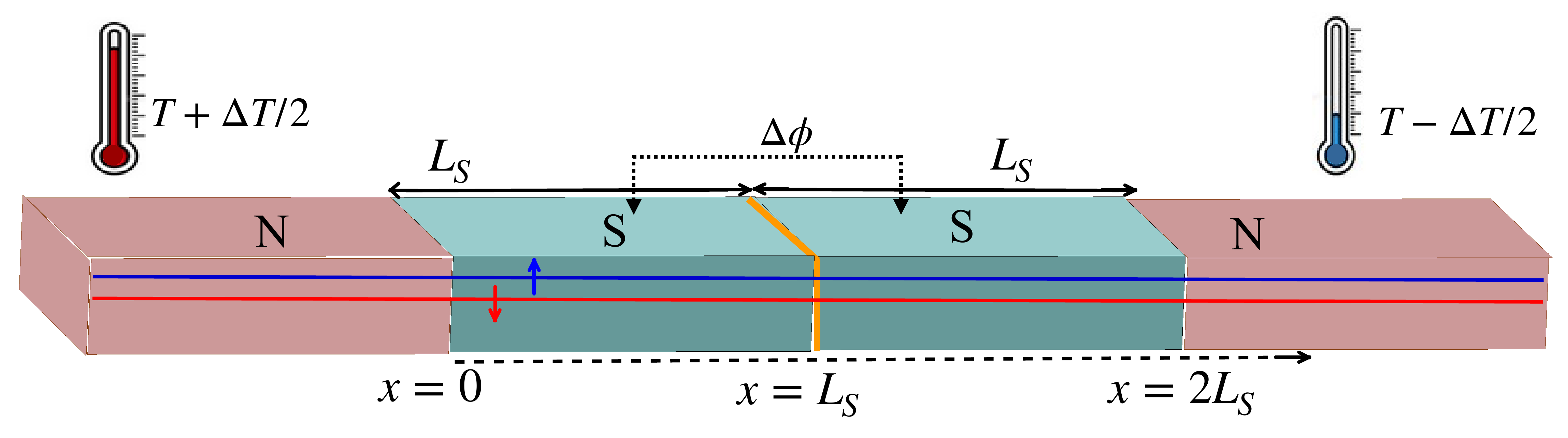}
\caption{N-sJJ-N junction with two ends maintained at two different temperatures. The whole junction is placed at the $1$D helical edge states shown by the red and blue lines with counterpropagating channels for up and down spin marked by blue and red arrow, respectively.}
\label{fig:model}
\end{figure}
%%%%%%%%%%%%%%%%%%%%%%%%%%%%%%%%%%%%%%%%%%%%%%%%%  

We consider a sJJ where two finite size superconductors, each having length $L_{\rm S}$, are coupled via a tiny insulator region. We take this insulator region as tiny just to simplify the calculation. A finite width of the insulator region will not affect our results qualitatively. The junction is formed at the edge of a $2$D\,TI and attached to two normal metal regions on opposite sides to form N-sJJ-N set-up. The superconductivity is proximity induced by using a traditional BCS superconductor as presented in Fig.\,\ref{fig:model}. The lengths of the two superconductors are set exactly equal to each other (denoted by $L_{\rm S}$) for simplicity. A small difference between them will not affect our results qualitatively. The phase difference between the two superconductors of the sJJ can be tuned by external magnetic flux $\phi$. We describe each part of the N-sJJ-N junction by Bogoliubov-de Gennes (BdG) Hamiltonian in the basis 
$\Psi (x) =\left(
 \psi_{\uparrow} (x),
 \psi_{\downarrow}(x),
\psi_{\downarrow}^{\dagger}(x),
-\psi_{\uparrow}^{\dagger}(x)\right)$ as\,\cite{Cayao2017},
\beq
\mathcal{H}_{\rm{BdG}}=
\begin{pmatrix}
H & \tilde{\Delta}\\
\tilde{\Delta}^\dagger& -H_{\text{}}^{\dagger}
\end{pmatrix},
\label{Eq:BdG_H}
\eeq
where the normal part Hamiltonian is given by
\bea
H=-i v_{F}\partial_{x}\sigma_{z}-\mu\sigma_{0}.
\label{Eq:h_normal}
\eea
The first term of Eq.\,\eqref{Eq:h_normal} is the kinetic energy term following the linear dispersion relation of the $1$D metallic edge states of $2$D\,TI. The second term includes the chemical potential $\mu$. The Pauli matrices $\sigma_{i}$, act in spin space and $\psi^{\dagger}_{\sigma}(x)$ ($\psi_{\sigma}(x)$) is the creation (annihilation) operator for an electron with spin $\sigma\!\in\!\{\uparrow,\downarrow\}$ at position $x$. The off-diagonal matrices of Eq.\,\eqref{Eq:BdG_H} are responsible for the  proximity-induced superconductivity described by the pair potential as: $\tilde{\Delta}(x)\!=\!\Delta(x) i \sigma_y$. We set it as: $\Delta(x)\!=\!\Delta_l$ for the left ($0 < x < L_{\rm S}$) and $\Delta(x)\!=\!\Delta_r e^{i \phi}$ for the right superconductor ($L_{\rm S} < x < 2L_{\rm S}$)  to have a finite phase difference in our sJJ, otherwise $\Delta(x)\!=\!0$ in all normal regions. We set the Fermi velocity $v_F\!=\!1$ and $\Delta_0\!=\!1$ so that for the symmetric junction where $\Delta_{\rm L}\!=\!\Delta_{\rm R}\!=\!\Delta$, the superconducting coherence length is $\xi\!=\!\hbar v_F/\Delta\!$\,. We show all the results for $\mu_N\!=\!0$ (for normal regions) and $\mu_S\!=\!2$ (for superconducting regions) but our results are insensitive to the chemical potential qualitatively.

The temperature dependence of the superconducting gap is taken following the relation $\Delta(T)\!=\!\Delta_0 \text{Tanh}(1.74 \sqrt{T_c/T - 1})$ where $T$ is the system temperature. For the symmetric junction, we take $\Delta_{\rm L}\!=\!\Delta_{\rm R}$ and show all the results for $T/T_c=0.3$. On the other hand, for the asymmetric junction, we consider $\Delta_{\rm L}\!\ne\!\Delta_{\rm R}$. To understand the effect of the gap asymmetry on the transport properties clearly, we maximize the difference between two gaps ($\Delta_L\!-\!\Delta_R$). To model the asymmetry, we take $\Delta_{\rm L}=\Delta(0.7)$ \ie reduced superconducting gap corresponding to $T/T_c=0.7$ and $\Delta_{\rm R}\!=\!\Delta_0$.

\section{Theoretical formalism}

We consider a temperature gradient across the junction without any bias voltage. The temperatures of the two leads are maintained at $T+\Delta T/2$ and $T-\Delta T/2$ (as shown in Fig.\,\ref{fig:model}) to set the temperature difference across the junction as $\Delta T$. Note that, $T$ is scaled by the superconducting transition temperature $T_c$. The applied temperature gradient acts in two ways: (i) it tunes the gaps in the density of states of the two superconductors of the sJJ, and (ii) it also affects the quasiparticles' occupation factors in the junction\,\cite{Pershoguba2019}. Consequently, there appear two different types of currents: charge current and heat current. The variation in superconducting gaps affects the usual Josephson current, which is non-dissipative. It can be expressed in terms of the variation of the gap as: $\delta I_c=\sum_l(\partial I_c/\partial \Delta_l)\delta \Delta_l$ considering the contribution by each lead $l$ connected to the superconductor with gap $\Delta_l$\,\cite{Pershoguba2019}. On the other hand, the occupation factor affects the charge current induced by the thermal gradient, which is dissipative. The dissipative and non-dissipative parts of the charge current can be separated by reversing the sign of the superconducting phase $\phi$. The dissipative part is even in phase $\phi$ \ie $I(\phi)\!=\!I({-\phi})$, whereas the non-dissipative component is odd in $\phi$: $I(\phi)\!=-\!I(\phi)$\,\cite{Pershoguba2019}.

{\it Charge current}: To evaluate the charge current induced by the temperature gradient $\Delta T$, we employ the Landauer transport theory. It can be written as the difference between the currents flowing in the opposite directions (coming from opposite leads) as\,\cite{Dutta2017,Pershoguba2019} $I^c= I_{\rm L}^c - I_{\rm R}^c$ where
\begin{align}
I^c_l &= \frac{2e}{h} \int_0^\infty d\omega \left[  i^{e}_l(\omega) - i^{h}_l(\omega) \right] f\left(\omega/T_l\right),
\label{Ic_LR}
\end{align}
where $e$ is the electronic charge, $h$ is the Planck's constant, $\omega$ is the incoming electron energy, and $f$ is the Fermi distribution function. Here, $l$ stands for L or R to represent the left or right normal metal leads, respectively, and $i^{e(h)}_l$ denotes the contributions by the electrons (holes) in $l$-th lead accordingly. Now, the heat current should be zero for $\Delta T=0$ following the second law of thermodynamics. Assuming this, the net current due to the temperature gradient can be calculated in terms of only one lead as
\begin{align}
I^c &= \frac{2e}{h} \Delta T \int_0^\infty d\omega \left[  i^{e}_{\rm L}(\omega)  -  i^{h}_{\rm L}(\omega) \right] \frac{\partial f\left(\omega/T_l\right)}{\partial T}. 
\label{Ic_l}
\end{align}
Note that, a finite amount of usual non-dissipative Josephson current is always present in the system even at $\Delta T=0$ for $\Delta \phi\ne 0$. The conservation of charge current due to the condensate can be taken care by performing fully self-consistent calculation\,\cite{Sanchez1997,Sanchez1998}. Since, we are only interested in the temperature driven part (\ie $\Delta T\ne0 $), we calculate the current in terms of one lead using Eq.\,\eqref{Ic_l}. The lower limit of the integration in Eq.\,\eqref{Ic_l} is to be replaced by the maximum among $\Delta_L$ and $\Delta_R$ if $T_{\rm ee}^{\rm R}\!=0$ within the subgap energy.
%%%%%%%%%%%%%%%%%%%% FIGURE OF Tee for symmetric %%%%%%
\begin{figure}[!thpb]
\centering
%$~~~~~\phi=\phi_0/4~~~~~~~~~~~$ $~~~~~~~~~~~~~~~~~~\phi=3\phi_0/4$\\
\includegraphics[scale=0.55]{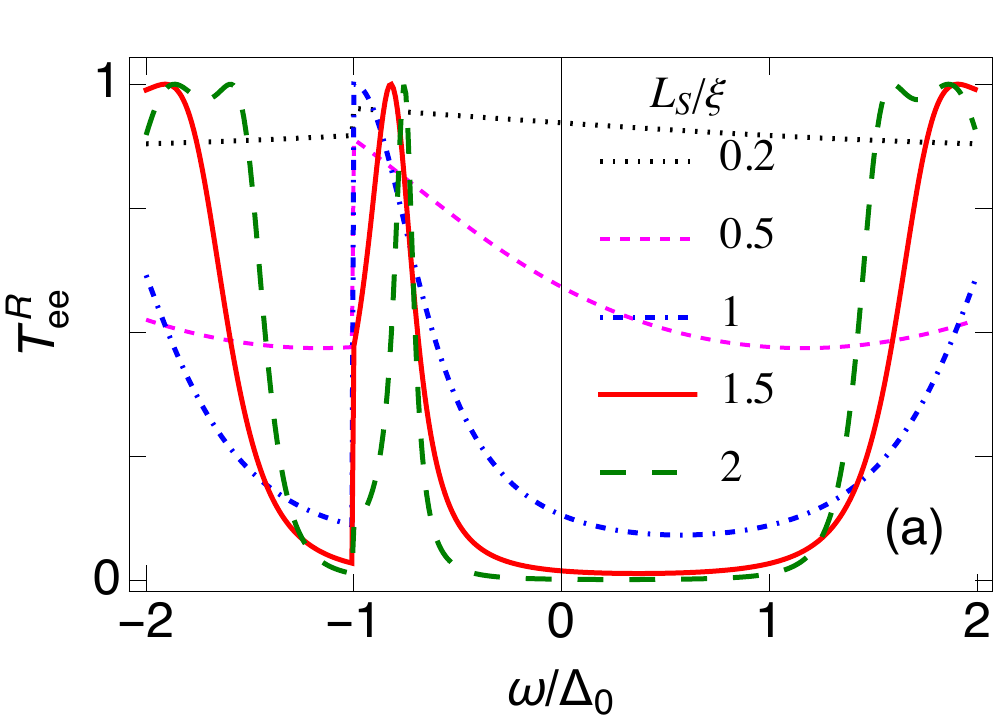}
\includegraphics[scale=0.55]{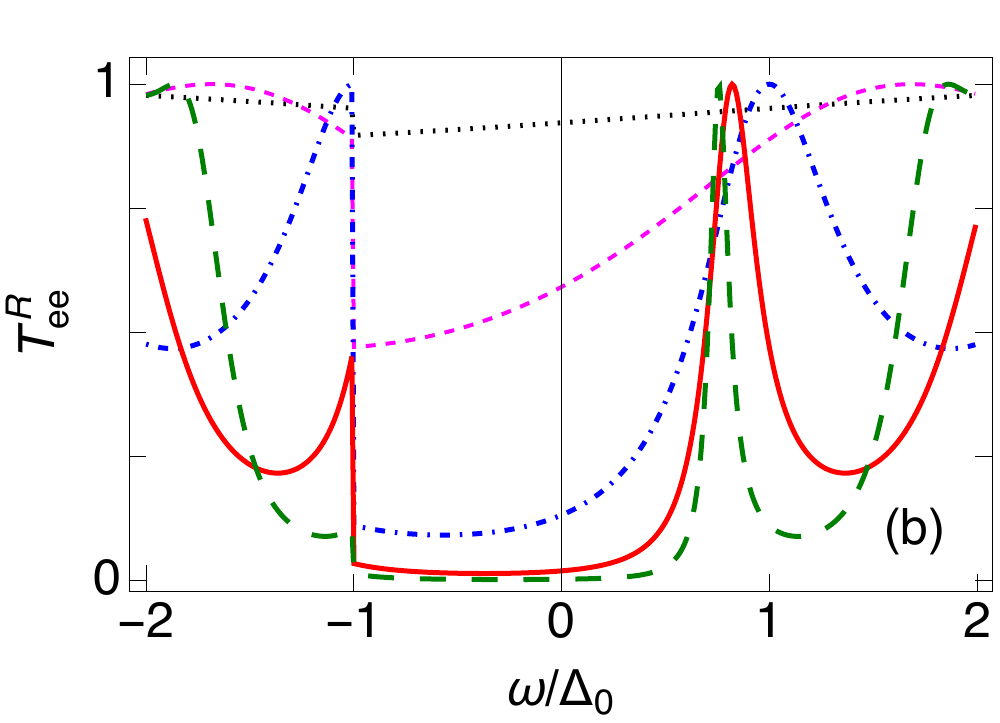}
\caption{Transmission probability $T_{\rm ee}^{\rm R}$ as a function of $\omega/\Delta_0$ in symmetric junction for (a) $\phi=\phi_0/4$ and (b) $\phi=3\phi_0/4$ with $\phi_0=2\pi$.}
\label{fig:ampl_sym}
\end{figure}
%%%%%%%%%%%%%%%%%%%%%%%%%%%%%%%%%%%%%%%%%%%%%%%%%%%%%%

Following the current conservation, the charge current should be continuous and we can find it out using the BdG wavefunctions and finally express it in terms of the transmission probabilities given by,
\begin{align}
 i_L^\eta=T_{\eta\eta}^{RL} - T_{\eta^{\prime}\eta}^{RL}
 \label{eq:ie}
\end{align}
with $\eta \in {\rm \{e, h\}}$ and $T_{\eta\eta}^{l^{\prime}l}=| t_{\eta\eta}^{l^{\prime}l} |^2$ where $T^{l^{\prime}l}_{\eta^{\prime}\eta}$ ($t^{l^{\prime}l}_{\eta^{\prime}\eta}$) is the probability (amplitude) of the transmission of $\eta^{\prime}$ type particles from $l^{\prime}$-th to $l$-th lead as $\eta$. In our case, $T_{\eta^{\prime}\eta}^{l^{\prime}l}=0$ when $l\ne l^{\prime}$ for $\eta\ne \eta^{\prime}$. The quasiparticles' transmissions take part in the dissipative part of the thermally induced charge current. The expressions for the transmission amplitudes are mentioned in the Appendix\,\ref{apnd1:TR}. 
%Note that, we neglect the terms proportional to $\omega/E_F$. 
From now on, we will use the notation $T_{\eta\eta}^{R}$ in place of  $T_{\eta\eta}^{RL}$ throughout the rest of the manuscript for simplicity.

Now, in the absence of any bias voltage, the linear response of the non-dissipative charge current per unit temperature difference is denoted as
\begin{equation}
\mathcal{L}_{12}=\frac{I^c}{\Delta T}.
\label{eq:L12}
\end{equation}
Note that, this is not the conventional Seebeck current as we explain in the next section. Reversing the phase can help in separating the non-dissipative charge current from the dissipative Seebeck current\,\cite{Pethick1979,Clarke1979,Pershoguba2019}.
%%%%%%%%%%%% FIGURE L12 for Symmetric junction %%%%%%%%
\begin{figure*}[!thpb]
\centering
\includegraphics[scale=0.57]{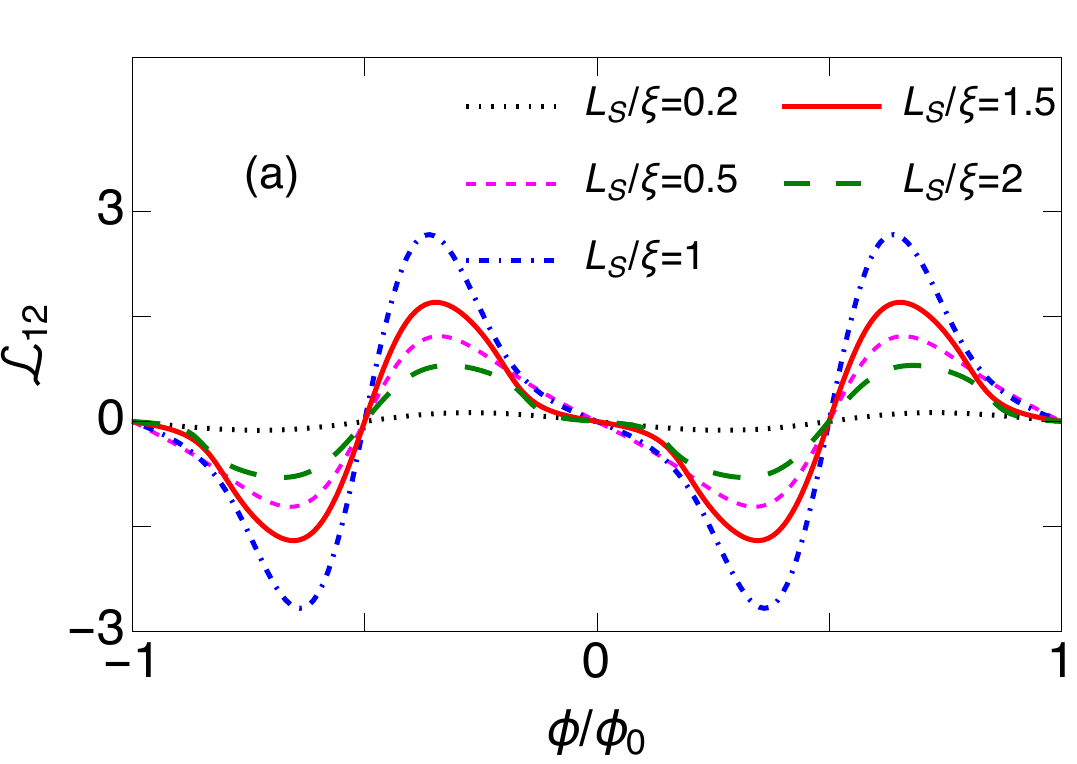}~~~~~~
\includegraphics[scale=0.4]{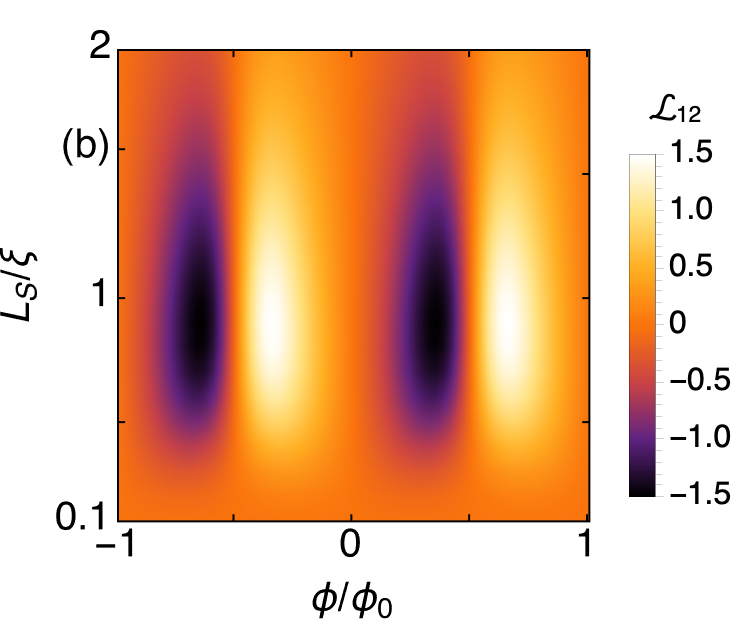}
\caption{Non-dissipative charge current per unit temperature gradient $\mathcal{L}_{12}$ (in units of $ek_B/h$) for symmetric junction as a function of (a) $\phi/\phi_0$ and (b) $\phi/\phi_0$ \& $L_{\rm S}/\xi$ with $\phi_0=2\pi$.}
\label{fig:Ssym}
\end{figure*}
%%%%%%%%%%%%%%%%%%%%%%%%%%%%%%%%%%%%%%%%%%%%%%%%%%%%%%

{\it Heat current}: To calculate the heat current, we follow the similar prescription considering the contributions by the individual leads as $I^q=I^q_L-I^q_R$ where 
\begin{align}
I^q_l =\int_0^\infty \omega d\omega 
\left[ i^e_l(\omega) + i^{h}_l(\omega) \right] f\left(\omega/T_l\right). 
\label{eq:Iq_l}
\end{align}
Using the initial condition that the heat currents flowing in the opposite direction must cancel each other for $\Delta T\!=\!0$, we finally arrive at
\begin{align}
I^q &= \frac{2}{h} \Delta T \int_0^\infty d\omega \left[  i^{e}_{\rm L}(\omega)  +  i^{h}_{\rm L}(\omega) \right] \frac{\partial f\left(\omega/T_l\right)}{\partial T}. 
\label{Iq}
\end{align}
where the contributions by the electron-like and hole-like quasiparticles are give by Eq.\eqref{eq:ie}. The 
heat current per unit temperature difference is defined as the thermal conductance and it is given by 
\beq
\mathcal{K}=\frac{I^q}{\Delta T}.
\eeq
We calculate charge current per unit temperature gradient $\mathcal{L}_{12}$ (in units of $ek_B/h$) and thermal conductance $\mathcal{K}$ (in units of $k_B/h$) for our sJJ considering small temperature gradient \ie $\Delta T\ll T/T_c$ within the linear response regime.

%%%%%%%%%%%%%%%%%%%%%%%%%%%%%%%%%%%%%%%%%%%%%%%%%%%%%%
\section{Results and Discussions}
%%%%%%%%%%%%%%%%%%%%%%%%%%%%%%%%%%%%%%%%%%%%%%%%%%%%%%

We compute the charge and heat current and present the results in this section. For the sake of understanding of the behaviors of the currents, we also investigate the quasiparticle transmissions throughout our N-sJJ-N junction. Since the temperatures of the two normal regions are different due to the temperature gradient across the N-sJJ-N junction, it is expected that it will affect the nearby superconductors accordingly to have different superconducting gaps. To explore the effect of the gap asymmetry in detail, we discuss both junctions with symmetrical and asymmetrical superconducting gaps called symmetric and asymmetric junctions, respectively, in the following subsections.

%---------------------------------------------
\subsection{Symmetric junction ($\Delta_L\!=\!\Delta_R\!$)}
%---------------------------------------------

We start by considering the simplest scenario where both the superconductors of the sJJ have the same gaps determined by the system temperature $T/T_c$.

\subsubsection{Transmission probability}

In order to understand the behaviors of the charge and heat currents flowing through the N-sJJ-N junction, we analyze the behaviors of the transmission spectra at first. We employ the scattering matrix method to calculate the transmission probability and present them in Fig.\,\ref{fig:ampl_sym}. For the details of the formalism and expression of the transmission probability, $T_{\rm ee}^{\rm R}$, we refer to Appendix\,\ref{apnd1:TR}.
%%%%%%%%%%%%%%% FIGURE OF K for symmetric junction %%%%%%%%
\begin{figure*}[!thpb]
\centering
\includegraphics[scale=0.57]{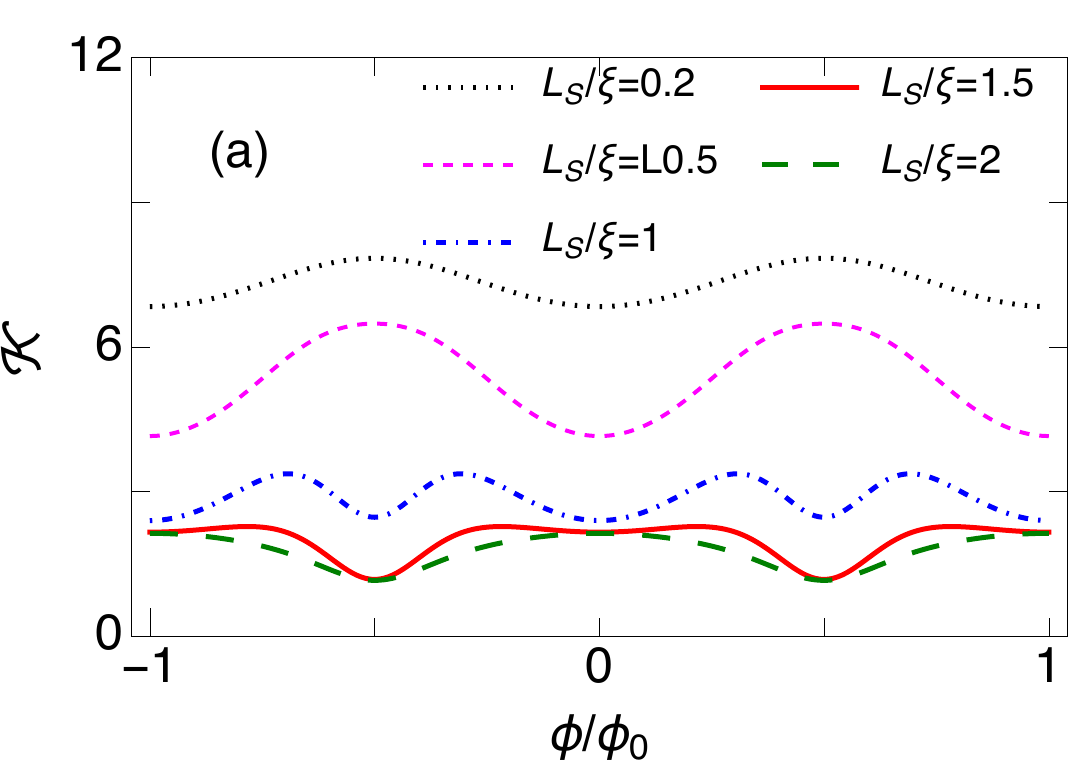}~~~~~
\includegraphics[scale=0.41]{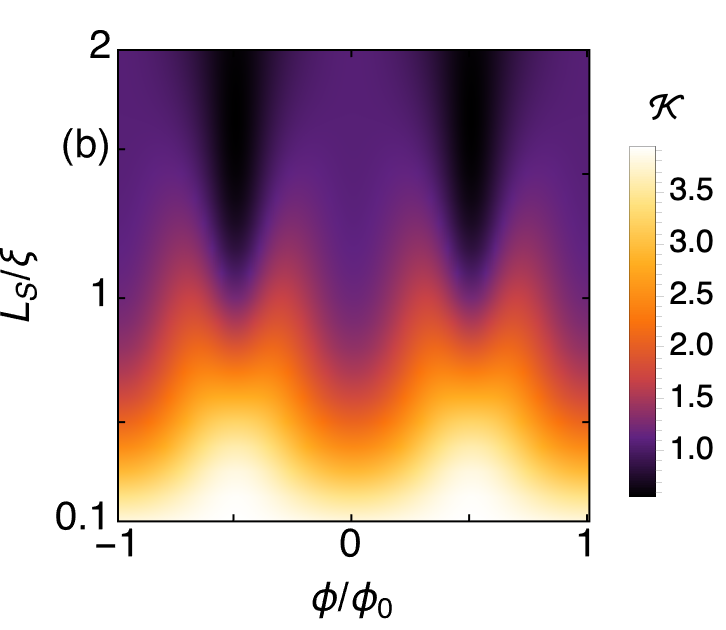}
\caption{Thermal conductance $\mathcal{K}$ (in units of $k_B/h$) for symmetric junction as a function of (a) $\phi/\phi_0$ and (b) $\phi/\phi_0$ (with $\phi_0=2\pi$) and $L_{\rm S}/\xi$.}
\label{fig:Ksym}
\end{figure*}
%%%%%%%%%%%%%%%%%%%%%%%%%%%%%%%%%%%%%%%%%%%%%%%%%%%%%%%

In Fig.\,\ref{fig:ampl_sym}, we show the results for two values of the phase difference ($\phi\!=\!\phi_0/4$ and $3\phi_0/4$ with $\phi_0=2\pi$) across the junction and various lengths $L_{\rm S}$ of the superconductors. We see that the spectra is asymmetric with respect to $\omega\!=\!0$ for both phases. This asymmetry exists as long as the phase is neither zero \ie $\phi\!\ne\!0$, nor half-integer multiples of $\phi_0$ \ie $\phi/\phi_0\ne n$ where $n$ is a half-integer and $\phi_0\!=\!2\pi$. In the absence of any phase difference between the two superconductors \ie $\phi\!=\!0$, the transmission amplitude is zero throughout the energy gap window with coherence peaks at the edges of the gap\,\cite{Cayao2022}, similar to what we get in any ordinary transparent normal metal/superconductor junction\,\cite{Paul2017}. Because of the helical nature of the edge states of the TI, there is no ordinary reflection to take place in the junction. The zero transmission is compensated by the unity Andreev reflection following the unitarity relation. However, the situation becomes dramatic when we tune the superconducting phase. By tuning the phase, the transmission peaks associated with a reduction in Andreev reflection (guaranteed by the unitarity relation between them) are found to exist due to the formation of ABS at the junction as discussed in Ref.\,[\onlinecite{Cayao2022}]. In the present work, we are only interested in other phases which are not discussed in Ref.\,[\onlinecite{Cayao2022}]. For any finite phase other than the time-reversal symmetric point set by $\phi/\phi_0\!=\!n$, the transmission peaks are asymmetrically positioned around $\omega\!=\!0$. The symmetry breaking around $\omega\!=\!0$ is true for any finite $L_{\rm S}$, but the transmission peaks get flattened with the decrease in the size of the two superconductors. Specifically, when $L_{\rm S}\!\ll\!\xi$, the transmission amplitude is close to unity throughout the energy window. It shows prominent peaks when $L_{\rm S}\sim \xi$ and the transmission peaks get more sharp when $L_{\rm S}>\xi$. Naively, for any particular $L_{\rm S}$, the behaviors of the spectra for a particular phase within the range $0 < \phi < \phi_0/2$ with $\phi_0=2\pi$, the peak position gets almost inverted to its mirror image with respect to $\omega\!=\!0$ when we tune the phase to another symmetrically chosen value within the range $\phi_0/2< \phi<\phi_0$ with $\phi_0=2\pi$. We explore the role of this asymmetry present in the transmission spectra about $\omega\!=\!0$ in inducing both charge and heat currents through the junction. We refer to Fig.\ref{fig:Tee_density_sym} for density plot of the transmission probability to check the results for other values of $L_{\rm S}$.

\subsubsection{Charge current}

With the understanding of the transmission probability, we present the results for charge current per unit temperature gradient, when the junction is only subject to the temperature gradient without any bias voltage. To see the effect of the symmetry breaking around $\omega\!=\!0$ for $\phi/\phi_0\ne n$ (where $n=0,1/2,3/2, ..$ and $\phi_0=2\pi$) on the charge current, we plot the charge current at $x=2L_{\rm S}$ in Eq.\,\eqref{eq:L12} as a function of $\phi/\phi_0$ and $L_{\rm S}$ in Fig.\,\ref{fig:Ssym}. Note that, this current denoted by $\mathcal{L}_{12}$ is non-dissipative part within the linear regime. 

In Fig.\,\ref{fig:Ssym}(a) we see that the behavior of the charge current is oscillatory with the change in $\phi/\phi_0$ with $\phi_0=2\pi$ maintaining zero amplitudes at $\phi/\phi_0\!=\!n$. There remains perfect symmetry around $\omega\!=\!0$ in the transmission spectra when $\phi/\phi_0\!=\!n$ as seen in the previous subsection. Tuning the phase difference to other finite values results in the symmetry breaking around $\omega\!=\!0$. The role of the symmetry can be confirmed from the nodes in the current profiles which exist for $\phi\!=\!0$ and $\phi/\phi_0\!=\!n$ with $\phi_0=2\pi$. The phase-tunable asymmetry in the transmissions of the quasiparticles within the subgap regime causes an imbalance between the left and rightly moving charges and as a consequence, a net charge current flows through the junction for all finite values of the phases except half-integer multiples of $\phi_0$. However, beyond the subgap limit, the transmission probabilities of the quasiparticles are finite for all phases of the sJJ.

The change in the peak positions, naively mirror inversion about $\omega\!=\!0$, by tuning the phase, in the transmission spectra reflects in the behavior of the charge current as well. It can be understood as follows. As soon as we tune the phase from zero to a positive finite value, the charge current starts increasing in amplitude (but with a negative sign) from zero and then again drops to zero at $\phi_0/2$ ($=\pi$). With further increase in $\phi$, the phase of the current reverses. The current amplitudes start increasing with positive sign and the profiles get inverted   when $\phi_0/2<\phi<\phi_0$ (equivalently, $\pi<\phi<2\pi$) compared to the profiles found for phases in the range $0<\phi<\phi_0/2$  (equivalently, $0<\phi<\pi$). This phase reversal corresponds to the naive mirror inversion of the transmission peaks around $\omega\!=\!0$ described in the previous subsection. 

The finite transmission of the quasiparticles via ABS in the subgap regime plays a major role in the charge current through the junction when biased by voltage difference as discussed in Ref.\,[\onlinecite{Cayao2022}]. The asymmetry around $\omega\!=\!0$ present in the transmission spectra plays a key role when the same system is driven by the thermal gradient. To investigate the role of the phase-tunable asymmetry, we provide some further results in Appendix\,\ref{apnd2:ABS}. Remarkably, the appearance of ABS is not accidental. It is protected by the topology of the $2$D\,TI and thus, enhances the possibility of utilizing our sJJ for the practical purpose. We also refer to Appendix\,\ref{apnd3:L12_T} for the discussions on the effect of the base temperature of the system. 

Next, we discuss the sensitivity of the charge current to the junction size. We observe that when $L_{\rm S}\ll \xi$, the amplitude of $\mathcal{L}_{12}$ increases with the increase in $L_{\rm S}$. In contrast, when $L_{\rm S}>\xi$, the behavior of the charge current changes. The current amplitude is decreasing with the rise in $L_{\rm S}$. To investigate the behavior of $\mathcal{L}_{12}$ in more detail, we present the density plot of $\mathcal{L}_{12}$ as a function of both $\phi/\phi_0$ (with $\phi_0=2\pi$) and $L_{\rm S}/\xi$ in Fig.\,\ref{fig:Ssym}(b). It is clear that the charge current amplitude is maximum when $L_{\rm S}/\xi \sim 1$. It shows opposite behavior, either increasing or decreasing with $L_{\rm S}$ in the two regimes defined by $L_{\rm S}/\xi\ll 1$ and $L_{\rm S}/\xi\gg 1$, respectively. Behavior of the charge current with $L_{\rm S}$ can also be explained by looking at the transmission spectra. When $L_{\rm S}\ll \xi$, there is almost uniform transmission throughout the subgap  regime. With the increase in $L_{\rm S}$, the asymmetry around $\omega\!=\!0$ starts to appear in the spectra, and that leads to increasing charge current with $L_{\rm S}$. On the other hand, the peak widths get much smaller when $L_{\rm S}/\xi\gg1$ resulting in decreasing behavior of the current through the junction.

As mentioned before, the dissipative and non-dissipative charge currents can be extracted from the total current by symmetrizing and anti-symmetrizing it with respect to the phase. We separate the even in $\phi$ and odd in $\phi$ part and confirm that in our model the dissipative part of the current is zero for the entire range of the phase difference. Hence, we show only the non-dissipative current which shows odd in $\phi$ symmetry. The whole charge current induced by temperature difference is non-dissipative current. Therefore, the symmetry breaking around $\omega$ can only generate the non-dissipative charge current in the short Josephson junctions within the linear regime. The reason behind the absence of dissipative charge current may be attribbuted to the interference between the currents carried by quasiparticles and Cooper pairs.

%-------------------------------------
\subsubsection{Heat current}
%-------------------------------------
In general, it is not compulsory to break the symmetry around $\omega=0$ to generate heat current. However, it is possible to tune the thermal current by introducing asymmetry in the junction. To see the effect of the symmetry breaking around $\omega\!=\!0$ on the heat current, we present the results of thermal conductance \ie heat current per unit temperature gradient as a function of $\phi/\phi_0$ in Fig.\,\ref{fig:Ksym}(a). 

We observe that similar to the charge current the behavior of the thermal conductance is also oscillatory with the phase of the junction. The heat current may have maximum values at $\phi/\phi_0\!=\!n$ with $n$ being either zero or half-integer and $\phi_0=2\pi$. This behavior is completely different from the behavior of the charge current. This happens because the energy carried by the electrons and holes are additive and they do not cancel with each other even when there exists symmetry around $\omega\!=\!0$ in the transmission spectra. However, the maxima of the heat current profiles for $L_{\rm S}/\xi\ll 1$ turns into minima when we increase the system size in the limit $L_{\rm S}/\xi\gg 1$. This can be explained by the presence of sharp peaks in the transmission probability spectrum. With the increase in the system size, the almost flat close to unity profiles change to have a few peaks and that effectively reduces the total transmission probability of the quasiparticles within the subgap regime and which further reduces the heat current. Also, unlike the behavior of the charge current, the behavior of the heat current is monotonic with $L_{\rm S}$. The heat current amplitude continuously decreases with the increase in $L_{\rm S}$. However, the rate of decrease of the heat current in the regime $L_{\rm S}/\xi \gg 1$ is much lower than the rate corresponding to the limit $L_{\rm S} \ll \xi$.
%%%%%%%%%%%%%%% FIGURE OF Tee for Asymmetric %%%%%%%%%%
\begin{figure}[!thpb]
\centering
%$~~~~~\phi=\phi_0/4~~~~~~~~~~~$ $~~~~~~~~~~~~~~~~~~\phi=3\phi_0/4$\\
\includegraphics[scale=0.55]{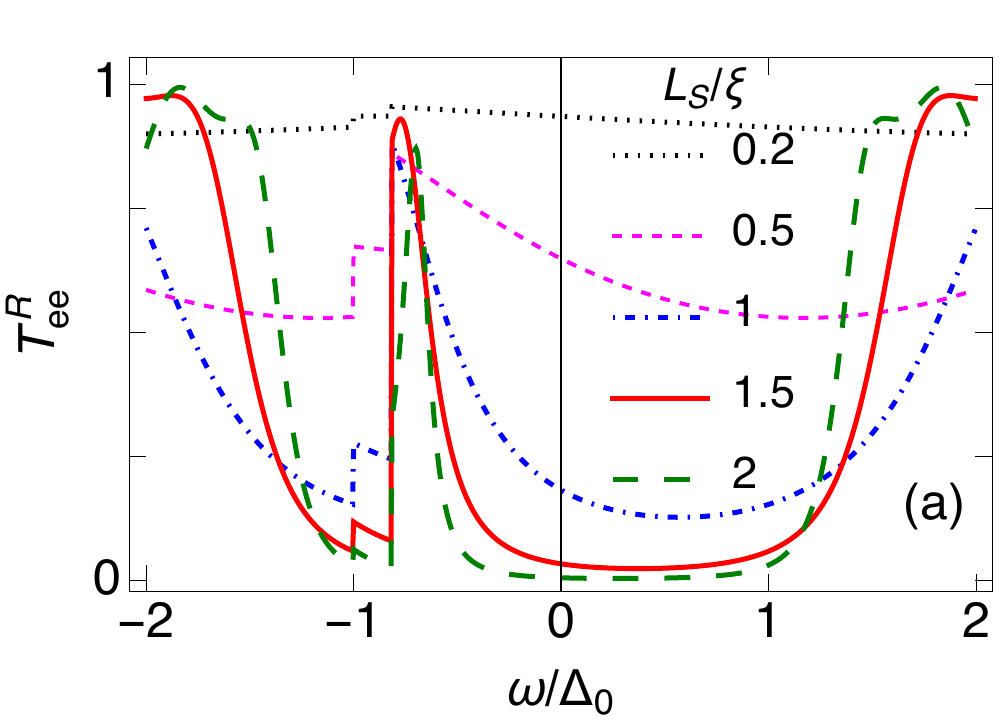}
\includegraphics[scale=0.55]{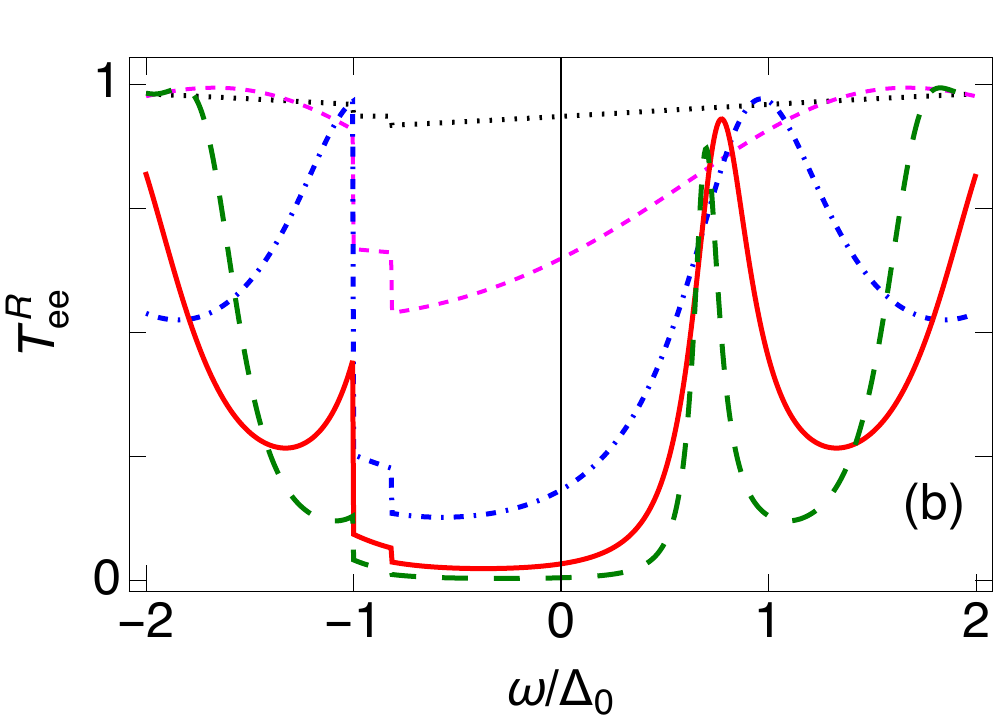}
\caption{Transmission probability $T_{\rm ee}^{\rm R}$ as a function of $\omega/\Delta_0$ in asymmetric junction for (a) $\phi=\phi_0/4$ and (b) $\phi=3\phi_0/4$ with $\phi_0=2\pi$.}
% with $\Delta_L\!=\!0.7\Delta$ and $\Delta_R\!=\!\Delta_0$.}
\label{fig:ampl_asym}
\end{figure}
%%%%%%%%%%%%%%%%%%%%%%%%%%%%%%%%%%%%%%%%%%%%%%%%%%%%%%

Note that, the heat current induced by the temperature gradient is higher for an extremely short junction. We reconfirm the same from the density plot of $\mathcal{K}$ as shown in Fig.\,\ref{fig:Ksym}(b). We see that the heat current amplitude is highest for the smallest size of the junction. It decreases when $L_{\rm S}/\xi \gtrsim 1$ being oscillatory with the phase across the junction. Similar to the charge current, the heat current is also phase-tunable. To increase the efficiency of any thermoelectric system, it is always recommended to minimize the thermal conductance and maximize the Seebeck coefficient\,\cite{Dutta2017,Dutta2019}. We can optimize this condition for our JJ in the limit $L_{\rm S}/\xi\sim 1$ and it is externally controllable by the phase of the junction. To understand the behaviors of $\mathcal{K}$ as a function of $T/T_c$, we refer to Appendix\,\ref{apnd2:ABS}.

%-------------------------------------------------------
\subsection{Asymmetric junction ($\Delta_L\! \ne\!\Delta_R\!$)}
%-------------------------------------------------------

Till now, our discussions are restricted to the symmetric junction where both the superconductors of our sJJ have similar gaps. In reality, as soon as we apply a temperature gradient between the normal regions, it is highly possible that the gaps between the two superconductors are modified accordingly since the superconductors are directly attached to the normal regions. To investigate the effect of this gap asymmetry on the thermal bias-induced current in sJJ, we present the results of transmission amplitudes, the charge and heat conductance for the condition of asymmetric superconducting gaps.
%%%%%%%%%%%%%%%%%%%% FIGURE OF K for asymmetric junction %%%
\begin{figure*}[!thpb]
\centering
\includegraphics[scale=0.63]{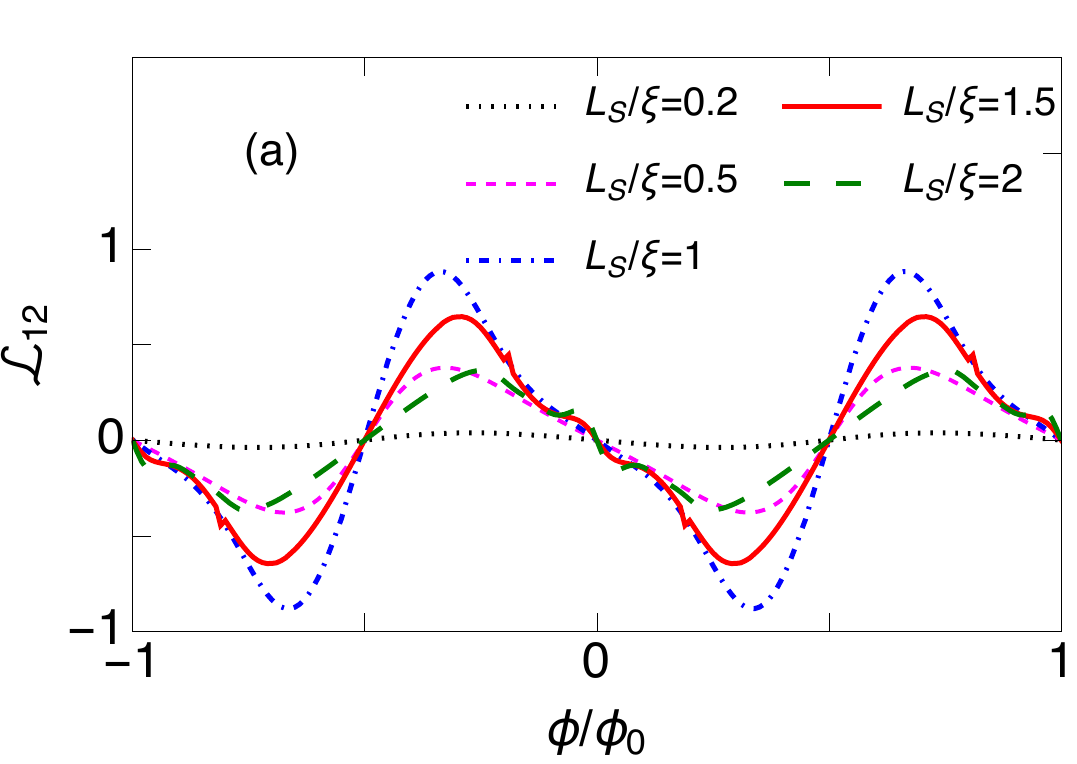}~~~~~~~~~
\includegraphics[scale=0.38]{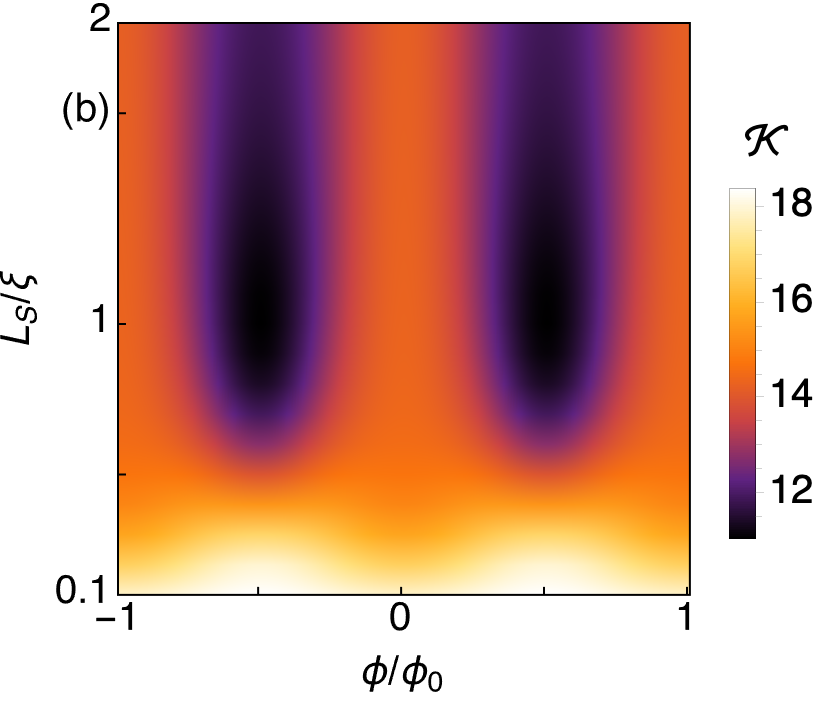}
\caption{Non-dissipative charge current per unit temperature gradient $\mathcal{L}_{12}$ (in units of $ek_B/h$) and thermal conductance $\mathcal{K}$ (in units of $k_B/h$) at $T/T_c\!=\!0.7$ for asymmetric junction as functions of $\phi$ and $L_S/\xi$.}
\label{fig:SKasym}
\end{figure*}
%%%%%%%%%%%%%%%%%%%%%%%%%%%%%%%%%%%%%%%%%%%%%%%%%%%%%
 
We refer to Fig.\,\ref{fig:ampl_asym} for the results of the probability of transmission of the quasiparticles in the asymmetric junction. We notice that in the transmission spectra, some new kinks appear when we take the gaps of the two superconductors different. The kinks correspond to the increased asymmetry in the system. For $L_{\rm S}/\xi \ll 1$, it gets more flattened with some additional kinks. For other limits of the lengths, the heights or widths of the peaks get reduced. Note that, we show the results for two different values of the phase difference of the sJJ. The qualitative behaviors of the transmission amplitudes, particularly the asymmetry around $\omega=0$ remain similar for all finite values except $\phi/\phi_0=n$. 

Now, we discuss the behaviors of the charge and heat current induced by temperature difference for the asymmetric junction as shown in Fig.\,\ref{fig:SKasym}. We observe that the non-dissipative charge current amplitude decreases for all limits of the lengths. The behavior of the charge current with the length and phase is similar to that in symmetric junction. This behavior of the charge current can be explained in terms of the transmission probability following the similar prescription as mentioned for symmetric junction. In contrast to the charge current, the heat current amplitude shows different behavior in the asymmetric junction. It shows a smaller magnitude for lower superconductor size compared to that in a symmetric junction. This will help in optimizing the current amplitudes for our sJJ in the limit $L_{\rm S}/\xi \sim 1$.

%-----------------------------------------------------------%
\section{Summary and Conclusions}\label{conclu}
%---------------------------------------------------------%
To summarize, we have explored the charge and heat currents flowing through a short JJ junction formed at the edge of $2$D\,TI. In order to understand the behavior of the currents, we have investigated the transmission probability as well. The asymmetry in the transmission spectra around zero energy, achieved by tuning the phase difference of the sJJ, induces the charge current and heat current through the junction. The phase-dependent currents are sensitive to the lengths of two finite sized superconductors of the junction. To optimize the currents in the sJJ, we recommend considering two superconductors of our short junction in the regime $L_{\rm S}/\xi\sim 1$. Note that, we have taken only the electronic contributions into account neglecting the phonon part for both symmetric and asymmetric junctions. It is justified since we are in the low-temperature regime. All charge currents we have shown, are entirely non-dissipative and show odd-symmetry in phase $\phi$. The symmetry breaking around $\omega=0$ can only generate non-dissipative part and is not sufficient for the generation of dissipative part of the charge current. We have only considered a part of the entire edge states. Adding contributions from the other part of the edge states will change our results quantitatively. However, the main message of the present study will remain invariant.

For the realization of the TI based sJJ, HgCd/HgTe and InAs/GaSb are good candidates for the TI as shown in Refs.\,[\onlinecite{Hart2014,Wiedenmann2016}]. For the proximity-induced superconductivity, any ordinary BCS superconductor e.g., Nb ($T_c\sim 9.2$K) can be used. Usage of high-temperature superconductors such as cuprate and iron-based superconductors will be helpful to have wide range for the base temperature. However, the temperature gradient is to be maintained low to validate the linear regime. Futher investigation can be performed to include non-linear regime and also to include the effect of different pairing mechanism in a separate report. Our anticipation for the existence of phase-dependent charge and heat current in sJJ and conclusions that {\it for short junction, the phase-difference is not sufficient to induce a dissipative charge current} thus providing valuable inputs to the discussions on the possibility of generating phase-induced thermal current in JJ.

%-------------------------------------------------------
\begin{acknowledgments}
PD acknowledges Jorge Cayao and Pablo Burset for helpful discussions and Department of Science and Technology (DST), India for the financial support through SERB Start-up Research Grant (File no.\,SRG/2022/001121).
\end{acknowledgments}
%--------------------------------------------------------

\onecolumngrid
\appendix

\section{Calculation of scattering amplitudes: scattering matrix formalism}\label{apnd1:TR}
In this Appendix, we describe the scattering matrix formalism which is employed to find the scattering amplitudes for our N-sJJ-N junction. The general form of the scattering states at the different
regions of the N-sJJ-N junction is given below. For symmetric junction, expressions for the scattering amplitudes drop down to the expressions mentioned in Ref.[\onlinecite{Cayao2022}].
%%%%%%%%%%%%%%%%%%%% FIGURE OF Tee for symmetric %%%%%%
\begin{figure}[!thpb]
\centering
%$~~~~~\phi=\phi_0/4~~~~~~~~~~~$ $~~~~~~~~~~~~~~~~~~\phi=3\phi_0/4$\\
\includegraphics[scale=0.48]{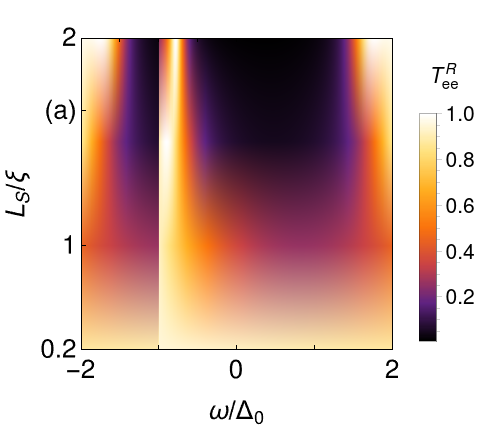}~~~~~~~~
\includegraphics[scale=0.48]{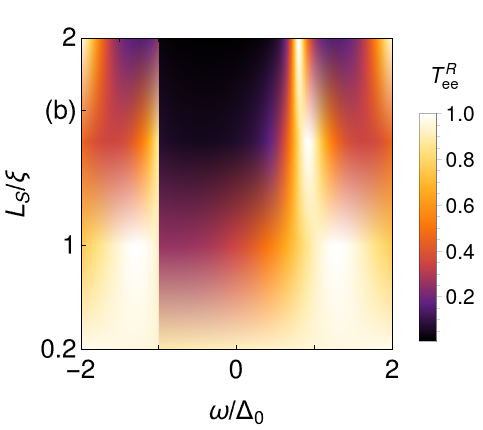}
\caption{Density plot of transmission probability $T_{\rm ee}^{\rm R}$ as a function of $\omega/\Delta_0$ and $L_{\rm S}/\xi$ in symmetric junction for (a) $\phi=\phi_0/4$ and (b) $\phi=3\phi_0/4$ with $\phi_0=2\pi$.}
\label{fig:Tee_density_sym}
\end{figure}
%%%%%%%%%%%%%%%%%%%%%%%%%%%%%%%%%%%%%%%%%%%%%%%%%%%%%%
\begin{widetext}
\bea
\Psi_{\rm 1}(x)&=&
 \psi_{1}^{N}\,{\rm e}^{ik_{e}x}+r_{\rm eh}^{\rm L}\psi_{3}^{N}\,{\rm e}^{ik_{h}x} +r_{\rm ee}^{\rm L}\psi_{2}^{N}\,{\rm e}^{-ik_{e}x}\,\quad x<0  \\
  &=&a_{1}\psi_{1}^{\rm S}\,{\rm e}^{ik^{\rm S_L}_{e}x}+b_{1}\psi_{2}^{\rm S_L}\,{\rm e}^{-ik^{\rm S_L}_{e}x}+c_{1}\psi_{3}^{\rm S_L}\,{\rm e}^{ik^{\rm S_L}_{h}x}+d_{1}\psi_{4}^{\rm S_L}\,{\rm e}^{-ik^{\rm S_L}_{h}x}~~{\rm for}~~ 0<x<L_{\rm S} \non \\
&=&   p_{1}\psi_{1}^{\rm S_R}\,{\rm e}^{ik^{\rm S_R}_{e}x}+q_{1}\psi_{2}^{\rm S_R}\,{\rm e}^{-ik^{\rm S_R}_{e}x}+r_{1}\psi_{3}^{\rm S_R}\,{\rm e}^{ik^{\rm S_R}_{h}x}+s_{1}\psi_{4}^{\rm S_R}\,{\rm e}^{-ik^{\rm S_R}_{h}x}~~{\rm for}~~ L_{\rm S}<x<2L_{\rm S} \non \\   
 &=& t_{\rm ee}^R  \psi_{1}^{\rm N}\,{\rm e}^{ik_{e}x} + t_{\rm eh}^{\rm R}  \psi_{4}^{\rm N}\,{\rm e}^{-ik_{h}x}~~{\rm for}~~x>2L_{\rm S} \non \\ 
\Psi_{2}(x)&=&
 \psi_{4}^{N}\,{\rm e}^{-ik_{h}x}+r_{\rm he}^{\rm L}\psi_{2}^{N}\,{\rm e}^{-ik_{e}x} +r_{\rm hh}^{\rm L} \psi_{3}^{N}\,{\rm e}^{ik_{h}x}~~{\rm for}~~x<0 \\
 &=& a_{2}\psi_{1}^{\rm S_L}\,{\rm e}^{ik^{\rm S_L}_{e}x}+ b_{2}\psi_{2}^{\rm S_L}\,{\rm e}^{-ik^{\rm S_L}_{e}x}+ c_{2}\psi_{3}^{\rm S_L}\,{\rm e}^{ik^{\rm S_L}_{h}x}+ d_{2}\psi_{4}^{\rm S_L}\,{\rm e}^{-ik^{\rm S_L}_{h}x}~~{\rm for}~~ 0<x<L_{\rm S} \non \\
  &=&p_{2}\psi_{1}^{\rm S_R}\,{\rm e}^{ik^{\rm S_R}_{e}x}+q_{2}\psi_{2}^{\rm S_R}\,{\rm e}^{-ik^{\rm S_R}_{e}x}+r_{2}\psi_{3}^{\rm S_R}\,{\rm e}^{ik^{\rm S_R}_{h}x}+s_{2}\psi_{4}^{\rm S_R}\,{\rm e}^{-ik^{\rm S_R}_{h}x}~~{\rm for}~~ L_{\rm S}<x<2L_{\rm S} \non \\ 
   &=& t_{\rm hh}^{\rm R}  \psi_{4}^{\rm N}\,{\rm e}^{-ik_{h}x} + t_{\rm he}^{\rm R}  \psi_{1}^{\rm N}\,{\rm e}^{ik_{e}x}~~{\rm for}~~\,x>2L_{\rm S} \non \\
\Psi_{3}(x)&=& c_{3}  \psi_{2}^{N}\,{\rm e}^{-ik_{e}x} +d_{3}  \psi_{3}^{N}\,{\rm e}^{ik_{h}x}~~{\rm for}~~x<0  \\
    &=&a_{3}\psi_{1}^{\rm S_L}\,{\rm e}^{ik^{\rm S_L}_{e}x}+b_{3}\psi_{2}^{\rm S_L}\,{\rm e}^{-ik^{\rm S_L}_{e}x}+c_{3}\psi_{3}^{\rm S_L}\,{\rm e}^{ik^{\rm S_L}_{h}x}+d_{3}\psi_{4}^{\rm S_L}\,{\rm e}^{-ik^{\rm S_L}_{h}x}~~{\rm for}~~ 0<x<L_{\rm S} \non \\
 &=& p_{3}\psi_{1}^{\rm S_R}\,{\rm e}^{ik^{\rm S_R}_{e}x}+q_{3}\psi_{2}^{\rm S_R}\,{\rm e}^{-ik^{\rm S_R}_{e}x}+r_{3}\psi_{3}^{\rm S_R}\,{\rm e}^{ik^{\rm S_R}_{h}x}+s_{3}\psi_{4}^{\rm S_R}\,{\rm e}^{-ik^{\rm S_R}_{h}x}~~{\rm for}~~ L_{\rm S}<x<2L_{\rm S} \non \\ 
   &=& \psi_{2}^{\rm N}\,{\rm e}^{-ik_{e}x}+a_{3}\psi_{4}^{\rm N}\,{\rm e}^{-ik_{h}x}+b_{3}\psi_{1}^{\rm N}\,{\rm e}^{ik_{e}x}~~{\rm for}~~x>2L_{\rm S}\non \\
\Psi_{4}(x)&=& c_{4}  \psi_{3}^{N}\,{\rm e}^{ik_{h}x} +    d_{4}  \psi_{2}^{N}\,{\rm e}^{-ik_{e}x}~~{\rm for}~~x<0   \\
    &=&a_{4}\psi_{1}^{\rm S_L}\,{\rm e}^{ik^{\rm S_L}_{e}x}+ b_{4}\psi_{2}^{\rm S_L}\,{\rm e}^{-ik^{\rm S_L}_{e}x}+ c_{4}\psi_{3}^{\rm S_L}\,{\rm e}^{ik^{\rm S_L}_{h}x}+ d_{4}\psi_{4}^{\rm S_L}\,{\rm e}^{-ik^{\rm S_L}_{h}x}~~{\rm for}~~0<x<L_{\rm S} \non \\
   &=&p_{4}\psi_{1}^{\rm S_R}\,{\rm e}^{ik^{\rm S_R}_{e}x}+ q_{4}\psi_{2}^{\rm S_R}\,{\rm e}^{-ik^{\rm S_R}_{e}x}+r_{4}\psi_{3}^{\rm S_R}\,{\rm e}^{ik^{\rm S_R}_{h}x}+s_{4}\psi_{4}^{\rm S_R}\,{\rm e}^{-ik^{\rm S_R}_{h}x}~~{\rm for}~~ L_{\rm S}<x<2L_{\rm S} \non \\ 
&=&\psi_{3}^{\rm N}\,{\rm e}^{ik_{h}x}+a_{4}\psi_{1}^{\rm N}\,{\rm e}^{ik_{e}x}+b_{4}\psi_{4}^{\rm N}\,{\rm e}^{-ik_{h}x}~~{\rm for}~~\,x>2L_{\rm S} \non
\label{phiNS}
\eea
\end{widetext}
where
\begin{widetext}
\begin{eqnarray}
\psi_{1}^{\rm N}\!&=&\!\begin{pmatrix} 1,0,0,0 \end{pmatrix}^{T},
\psi_{2}^{\rm N}\!=\!\begin{pmatrix} 0,1,0,0 \end{pmatrix}^{T},
\psi_{3}^{\rm N}\!=\!\begin{pmatrix} 0,0,1,0  \end{pmatrix}^{T},
\psi_{4}^{\rm N}\!=\!\begin{pmatrix} 0,0,0,1 \end{pmatrix}^{T}, \\ \non 
\psi_{1}^{\rm S_i}\!&=&\!\begin{pmatrix} u_i, 0,v_i,0  \end{pmatrix}^{T}, 
\psi_{2}^{\rm S_i}\!=\!\begin{pmatrix} 0,u_i,0,v_i \end{pmatrix}^{T},
\psi_{3}^{\rm S_i}\!=\!\begin{pmatrix} v_i,0,u_i,0  \end{pmatrix}^{T},
\psi_{4}^{\rm S_i}\!=\!\begin{pmatrix} 0,v_i,0,u_i  \end{pmatrix}^{T}.
\end{eqnarray}
\end{widetext}
Note that, here we use the notation N for normal regions, both left and right, since they have exactly similar parameters except for the gaps in the asymmetric case, as mentioned in the main text. The wave vectors in the normal regions are given by
\bea
k_{e(h)}(\omega)=\frac{\mu \pm \omega}{v_F}.
\eea
Within the superconductors the wave vectors take the form as
\bea
k_{e(h)}^{S_i}(\omega,\Delta_i)\!=\!\frac{\mu\pm \sqrt{\omega^2-\Delta^2_i}}{v_F}
\eea
with $i$ denoting L or R. The coherence factors for the two superconductors are given by 
\beq
 u_i,v_i={\bigg[\frac{\omega\pm\sqrt{\omega^{2}-\Delta_i^{2}}}{2\omega}}\bigg]^{1/2} .
 \eeq
 
 To solve the equations, the wave functions are matched at the three interfaces of the junction, at $x=0$, $x\!=\!L_{\rm S}$, and $x=2L_{\rm S}$. The ordinary reflection and crossed Andreev reflection are prevented due to the helicity of the edge states of $2$DTI\,\cite{Tkachov2013}, allowing only two processes: (1) Andreev reflections where an incident electron (a hole) is reflected as a hole (an electron) at the left normal region and (2) electron (hole) transmission at the right normal region. The corresponding transmission amplitudes are given as follow.
 
 For the transmission of electron in the right normal region after injecting an electron from the left normal region, the amplitude reads as
\begin{widetext}
\beq
t_{\rm RL}^{\rm ee}=\frac{(u_L^2-v_L^2)(u_R^2-v_R^2) e^{-2ik_{\rm e}+i(k_{\rm e}^{\rm S_L}+k_{\rm e}^{\rm S_R}+k_{\rm h}^{\rm S_L}+k_{\rm h}^{\rm S_R}) L_{\rm S}}~e^{i\phi}} {(e^{ik_{\rm e}^{\rm S_L} L_{\rm S}}-e^{ik_{\rm h}^{\rm S_L}L_{\rm S}})(e^{ik_{\rm e}^{\rm S_R} L_{\rm S}}-e^{ik_{\rm h}^{\rm S_R}L_{\rm S}}) u_L v_L u_Rv_R + (u_L^2 e^{ik_{\rm e}^{\rm S_L} L_{\rm S}}-v_L^2 e^{ik_{\rm h}^{\rm S_L}L_{\rm S}})(u_R^2e^{ik_{\rm e}^{\rm S_R} L_{\rm S}}-v_R^2e^{ik_{\rm h}^{\rm S_R}L_{\rm S}}) ~e^{i\phi}}.
\eeq
\end{widetext}
For the transmission of hole in the right normal region after injecting a hole from the left normal region, the amplitude is expressed as
\begin{widetext}
\beq
t^{\rm hh}_{\rm RL}=\frac{(u_L^2- v_L^2) (u_R^2- v_R^2) e^{2ik_{\rm h}L_{\rm S} } e^{-i\phi}}{ (e^{ik_{e}^{\rm S_L}L_{\rm S}}-{\rm e}^{ik_{\rm h}^{\rm S_L}L_{\rm S}})(e^{ik_{e}^{\rm S_R}L_{\rm S_R}}-{\rm e}^{ik_{\rm h}^{\rm S_R}L_{\rm S}})  u_L v_L u_Rv_R+(u_L^2 {\rm e}^{ik_{h}^{\rm S_L}L_{\rm S}}-v_L^2{\rm e}^{ik_{e}^{\rm S_L}L_{\rm S}})(u_R^2 {\rm e}^{ik_{h}^{\rm S_R}L_{\rm S}}-v_R^2{\rm e}^{ik_{e}^{\rm S_R}L_{\rm S}}) e^{-i\phi}}.
\eeq
\end{widetext}
It follows the unitarity relation: $R^{\rm eh}_{\rm LL}+T^{\rm hh}_{\rm RL}=1$ where $R^{\rm eh}_{\rm LL}=|r^{\rm eh}_{\rm LL}|^2$ and $T^{\rm hh}_{\rm RL}=|t^{\rm hh}_{\rm RL}|^2$.
Note that, the following relations hold for our sJJ formed at the helical states: $r_{\rm LL}^{\rm ee}\!=\!r_{\rm LL}^{\rm hh}\!=r_{\rm RR}^{\rm ee}\!=\!r_{\rm RR}^{\rm hh}\!=\!t_{\rm RL}^{\rm eh}\!=\!t_{\rm RL}^{\rm he}=\!t_{\rm LR}^{\rm eh}\!=\!t_{\rm LR}^{\rm he}\!=\!0$. 
 %%%%%%%%%%%%%%%%%%%% FIGURE for ABS in symmetric %%%%%%%%%%%
\begin{figure*}[!thpb]
\includegraphics[scale=0.65]{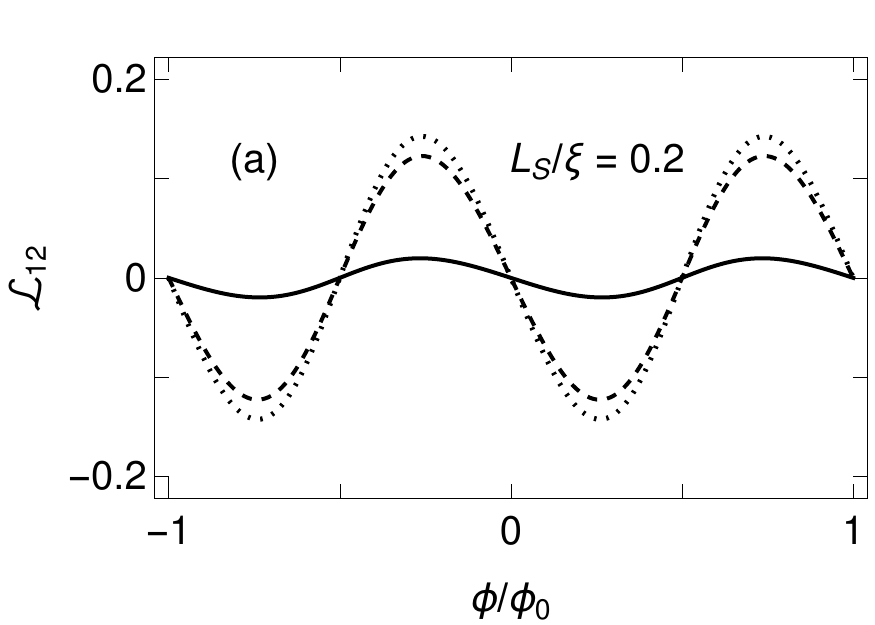}
\includegraphics[scale=0.65]{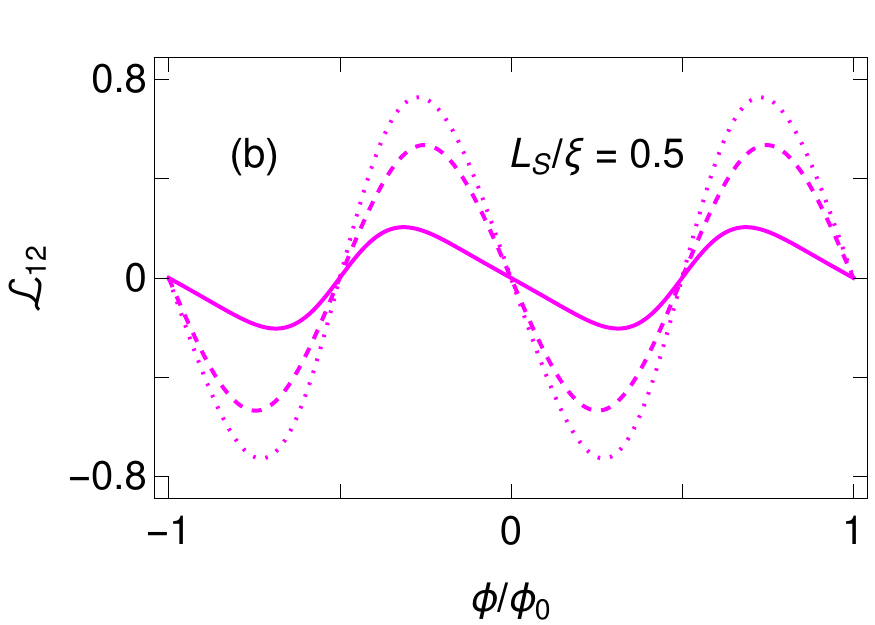} 
\includegraphics[scale=0.65]{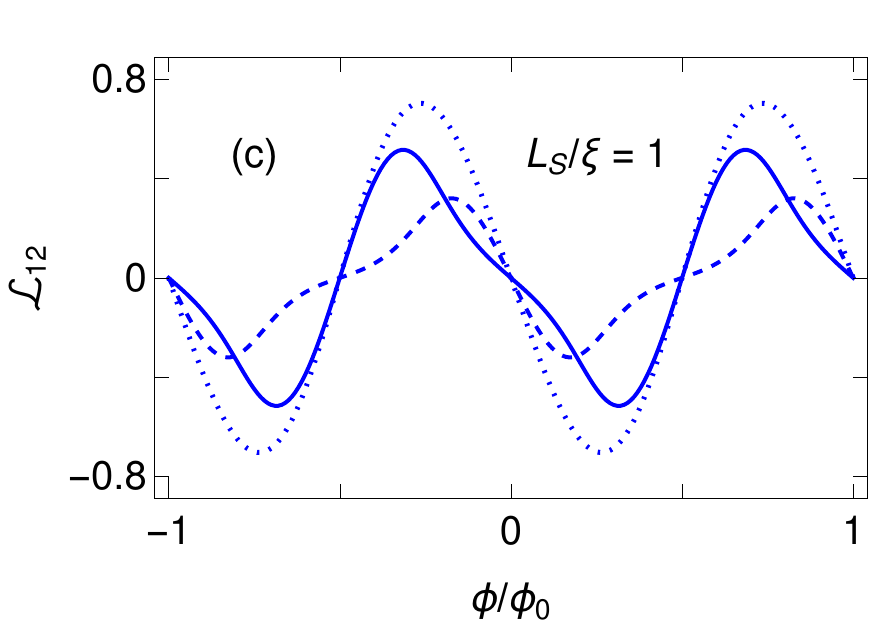} \\
\includegraphics[scale=0.65]{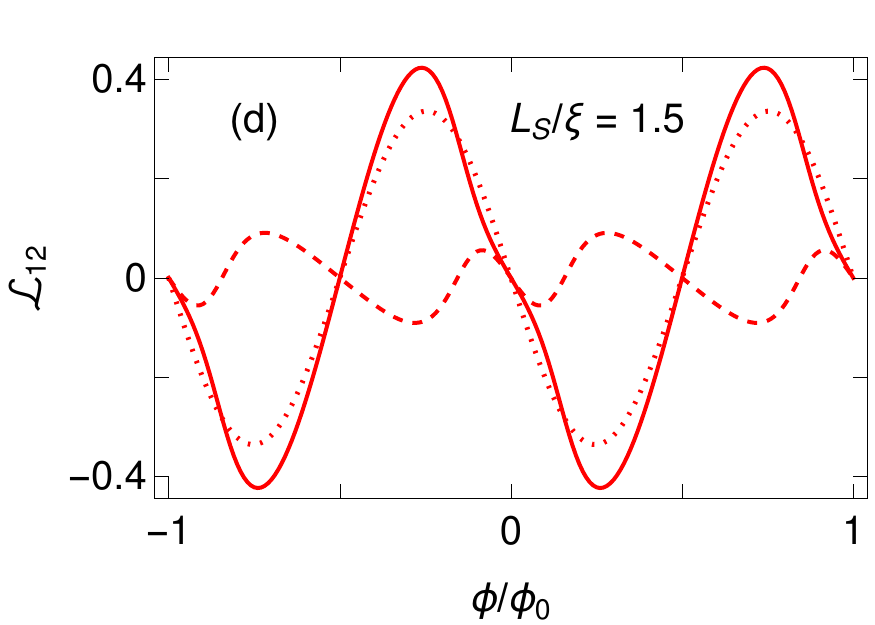}
\includegraphics[scale=0.65]{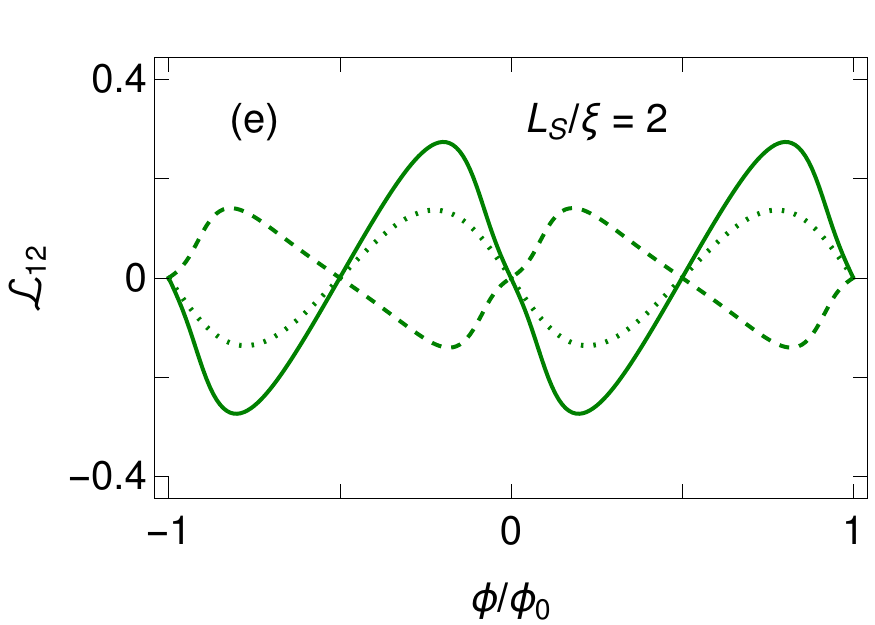}
\caption{Non-dissipative charge current per unit temperature gradient $\mathcal{L}_{12}$ (in units of $ek_B/h$) for symmetric junction as a function of $\phi/\phi_0$ for various $L_{\rm S}$. The solid, dashed and dotted lines represent the integration limits $[0,\Delta_0]$, $[\Delta_0,\infty]$, and $[0,\infty]$, respectively.}
\label{fig:L12_sym_Int} 
\end{figure*}
%%%%%%%%%%%%%%%%%%%%%%%%%%%%%%%%%%%%%%%%%%%%%%%%%

We plot the transmission probability for two phase values (as used in Fig.\,\ref{fig:ampl_sym} where some discrete values of $L_{\rm S}$ are considered) as a function of $\omega/\Delta_0$ and $L_{\rm S}/\xi$. We observe that the asymmetry in transmission amplitude with respect to $\omega=0$ increases with the increase in $L_{\rm S}$.

\section{The role of subgap asymmetry in charge current}\label{apnd2:ABS}
To unveil the role of the phase-tunable subgap asymmetry, we plot the charge current through the symmetric junction in Fig.\,\ref{fig:L12_sym_Int} by dividing the limit of the integration of Eq.\,\eqref{Ic_l} into two parts: $[0,\Delta_0]$, [$\Delta_0,\infty$]. The total current is found by setting the limit as $[0,\infty]$. This total charge current is the non-dissipative charge current as mentioned in the main text and in the following subsection. We show this charge current breakups for various sizes of the superconductors because of the sensitivity of the charge current to the superconductor size.

In Fig.\ref{fig:L12_sym_Int} we observe that for $L_{\rm S}/\xi\ll 1$, the total charge current in the junction is mostly dominated by the transmissions above the superconducting gap. The contributions by the subgap and supergap transmissions are in phase giving rise to the additive total charge current for an extremely short junction. However, the scenario changes when we increase the lengths of the two superconductors. The contributions by the subgap states increase with the increase of $L_{\rm S}$. When $L_{\rm S}/\xi\sim 1$, the major contribution to the charge current comes from the subgap asymmetry in the transmission spectra. Comparing all the sub-figures in Fig.\,\ref{fig:L12_sym_Int}, we observe that the total charge current is increasing with $L_{\rm S}$ initially but falls down when $L_{\rm S}/\xi> 1$. In fact, when $L_{\rm S}/\xi \gg 1$, the total charge current decreases and it is lower than the contributions by the current carried by the subgap states. 

It happens because of the phase change between the
contributions by the subgap and supergap contributions resulting in the enhancement of the suppression of the total charge current. For $L_{\rm S}/\xi\ll 1$, both subgap and super-gap contributions are in phase resulting into the higher total charge current. The scenario changes when $L_{\rm S}/\xi\sim 1$ and more prominently when $L_{\rm S}/\xi \gg 1$. The total current goes down as the contributions by the subgap and supergap states are completely out of phase but comparable in amplitudes. As a consequence, we see that the contributions by the subgap states to the thermally induced total charge current is highest and in phase when $L_{\rm S}/\xi$. Beyond this regime of $L_{\rm S}/\xi$, the contributions by ABS are majorly compensated by the contributions from the supergap states.

\section{Effects of temperature on charge current}\label{apnd3:L12_T}
In the main text, we have only presented the behaviors of the charge current at a particular temperature. In the present section, we discuss the behaviors of the charge current at different temperatures. Note that, unless we mention specifically, we always consider the whole range of the integration to calculate the charge current throughout the study. Fixing the base temperature of the system $T/T_c$ to various values, we present all the results by applying a small gradient around that base temperature. 
%%%%%%%%%%%%%%%%%%%% DESNITY plot of L12 with T for symmetric 
\begin{figure*}[!thpb]
$~~~~~~~{\rm(a)}~L_s/\xi=0.5~~~~~~~~~~~~~~~~~~~~~~~~~~~~~~~{\rm (b)}~L_s/\xi=1~~~~~~~~~~~~~~~~~~~~~~~~~~~~~~~~{\rm (c)}~L_s/\xi=1.5~~~~~~$\\
\includegraphics[scale=0.38]{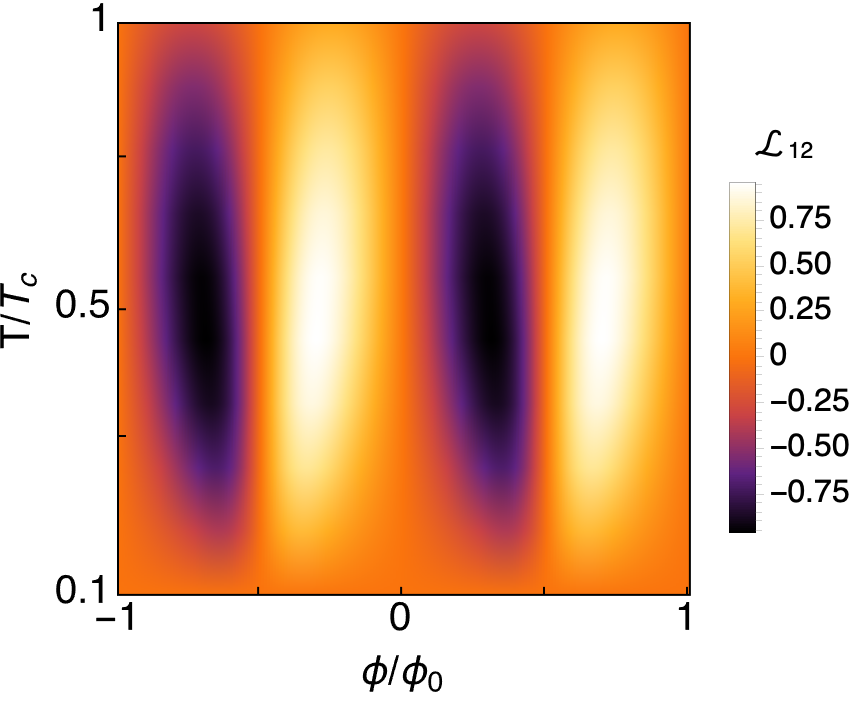}
\includegraphics[scale=0.38]{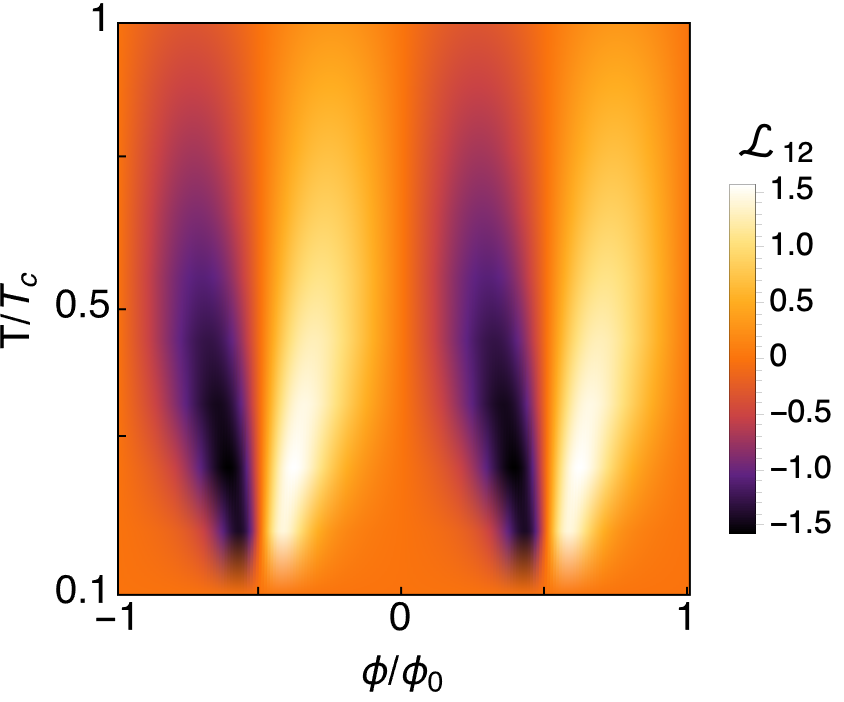}
\includegraphics[scale=0.38]{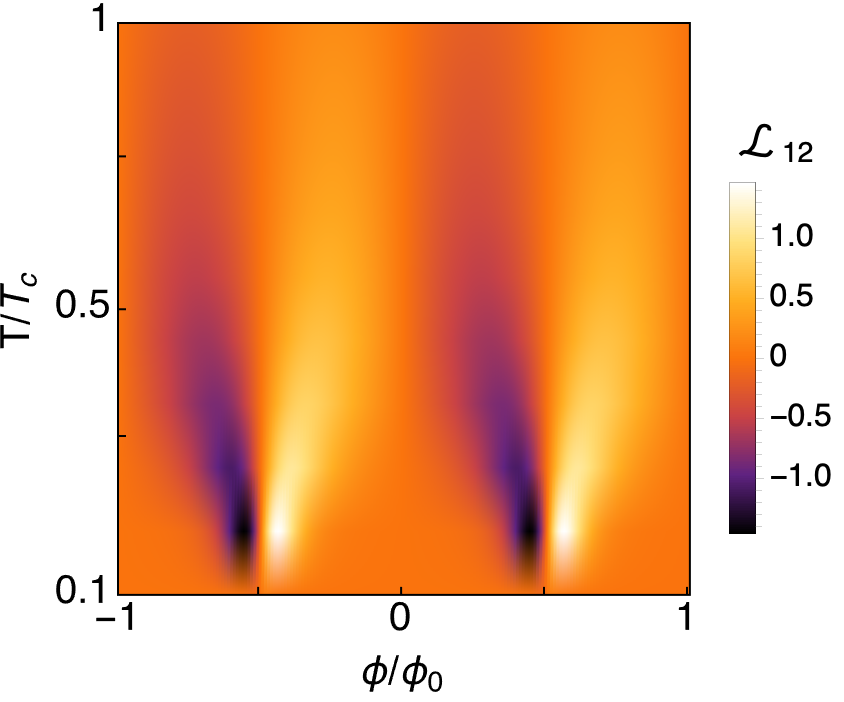}
\caption{Charge current per unit temperature gradient $\mathcal{L}_{12}$ (in units of $ek_B/h$) for symmetric junction as a function of $\phi$ and $T/T_c$ for various $L_{\rm S}/\xi$. }
%Limits of the Integration is set as [$0,\infty$].}
\label{fig:sym_T}
\end{figure*}
%%%%%%%%%%%%%%%%%%%%%%%%%%%%%%%%%%%%%%%%%%%%%%%%%%%%%%%

In Fig.\,\ref{fig:sym_T}, we show the density plots of non-dissiptaive charge current $\mathcal{L}_{12}$ as a function of temperature and phase difference of the junction. We see that the oscillatory behavior of the charge current as a function of $\phi/\phi_0$ as discussed in the main text. For $L_{\rm S}/\xi\ll 1$, the current amplitude becomes large when $T/T_c\sim 0.5$. At this regime of the superconductor length, the asymmetry around $\omega\!=\!0$ in the transmission spectra is much smaller. To enhance the current in this scenario, the system temperature has to be increased sufficiently. Increasing the temperature beyond this value will result in smaller gaps and thus reduces the ABS contributions to the current. This is confirmed when we take larger superconductors \ie $L_{\rm S}/\xi \sim 1$. At this limit, the asymmetry around $\omega\!=\!0$ in the transmission probability profile is much higher and we can get enhanced current even in the very low temperature limit. However, the picture changes when we take larger size superconductors in the limit $L_{\rm S}/\xi\gg 1$. At this limit of the size, the higher amplitude of the current is constrained to the very small regime of the temperature. This can be explained following the similar prescription mentioned for other limits of the size of the two superconductors.

\bibliography{bibfile}

\end{document}